\documentclass[twocolumn,aps,ams,epsfig,graphicx,showpacs,floatfix]{revtex4}
\usepackage{graphicx}
\newcommand{\be}{\begin{equation}}
\newcommand{\ee}{\end{equation}}

\newcommand{\bea}{\begin{eqnarray}}
\newcommand{\eea}{\end{eqnarray}}
\newcommand{\bd}{\begin{displaymath}}
\newcommand{\ed}{\end{displaymath}}
\newcommand{\bi}{\begin{itemize}}
\newcommand{\ei}{\end{itemize}}
\newcommand{\bc}{\begin{center}}
\newcommand{\ec}{\end{center}}
\newcommand{\bfl}{\begin{flushleft}}
\newcommand{\efl}{\end{flushleft}}
\newcommand{\bfr}{\begin{flushright}}
\newcommand{\efr}{\end{flushright}}
\newcommand{\f}{\frac}

\def\ra{\rightarrow}
\def\6{\partial}

\def\={\!\!\!&=&\!\!\!}
\def\+{\!\!\!&&\!\!\!+~}
\def\-{\!\!\!&&\!\!\!-~}

\begin{document}
\title{Zero temperature conductance of parallel T-shape double quantum dots}

\author{M. Crisan}
\author{I. Grosu}
\affiliation{Department of Physics, ``Babe\c{s}- Bolyai" University,
40084 Cluj-Napoca, Romania}
\author{I. \c{T}ifrea}
\affiliation{Department of Physics, California State University,
Fullerton, CA 92834, USA}

\begin{abstract}
We analyze the zero temperature conductance of a parallel T-shaped
double quantum dot system. We present an analytical expression for
the conductance of the system in terms of the total number of
electrons in both quantum dots. Our results confirm that the
system's conductance is strongly influenced by the dot which is not
directly connected to the leads. We discuss our results in
connection with similar results reported in the literature.
\end{abstract}
 \maketitle

\section{Introduction}
\begin{figure}[t]
\centering \scalebox{0.5}[0.5]{\includegraphics*{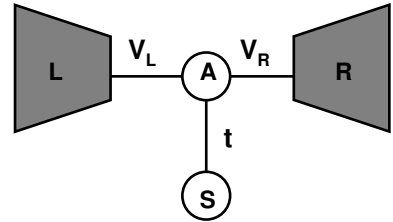}}
\caption{Schematic representation of the parallel T--shaped double
quantum dot system. The active dot A is connected to the left and
right leads and to the side dot S. The presence of the side dot S,
which is only connected to the active dot A, influences the general
conductance of the system.} \label{system}
\end{figure}

Recent advances in the fabrication and precise control of nanoscale
electronic systems lead to an increased interest in the study of
many body effects in quantum dot structures. The Anderson single
impurity model \cite{anderson} was extensively explored to
successfully understand electronic correlations in small single, or
double quantum dot structures. In general, single or double quantum
dot configurations provide the ideal systems to study many body
effects. For example, single dot configurations allow the
realization of the Kondo regime of the Anderson impurity
\cite{singledot}. On the other hand, double quantum dot (DQD)
configurations provide the ideal candidate for the study of the many
body effects associated to both Kondo effect and RKKY interaction
\cite{doubledots}. The connection of several quantum dots (QD) gives
rise to remarkable phenomena due to the interplay of electron
correlations and interference effects which depend on how the dots
are arranged. One possible configuration is the double quantum--dot
(DQD) system, where the dots are connected to the same leads and
between them. Recently, Dias da Silva {\em et al}. \cite{1} studied
a DQD with one dot in the Kondo regime and the other close to the
resonance with the connecting leads. One of the most interesting
results reported in Ref. \cite{1} is the finite temperature analysis
of the parallel T-shape double dot configuration with one dot
disconnected from the leads (Fig. \ref{system}) using the idea of
interference between resonances. In such a configuration, the active
dot A is directly connected to the left and right leads and to a
side dot S. They showed that when the side dot S is coupled to the
leads only through the active dot A, the Kondo resonance from the
side dot S develops a sizable splitting even if there is no magnetic
field in the system. This band filtering produced by the connected
dot preserves the Kondo singlet and at finite temperature the
magnetic moment is completely screened.

The calculation of system's conductance is of major interest both
for single and double quantum dots configurations. For double
quantum dot configurations the problem was considered extensively
using different methods, however, the results presented by different
authors are in agreement only partially \cite{kim,tak,co,zi}. The
system's electronic conductance is realized through the active dot
A, however, the presence of the side dot S will influence the total
conductance of the parallel T--shaped quantum dot system \cite{kim}.
The suppression of the system's conductance at low temperatures can
be understood if two possible conduction paths are considered, a
direct path through the active dot A ($L\ra A\ra R$) and an indirect
path through the side dot S ($L\ra A\ra S \ra A\ra R$). New features
have been pointed out by Takazawa {\em et al.} \cite{tak} in
connection with the inter--dot coupling strength $t$ and the values
of the energies of the active dot $E_A$ and the energy of the side
dot $E_S$ . For the case $E_A=E_S$ the occurrence of a non trivial
suppression of the conductance was associated with a Fano--like
effect between two distinct channels, i.e., the direct Kondo
resonance of the active dot A and the indirect resonance via both
the active and side dots. An interesting feature of the system was
discussed by Cornaglia and Grempel \cite{co} and by Zitko and Bonca
\cite{zi} using the new idea introduced in Ref. \cite{wiel} known as
the two-stage Kondo effect, a behavior obtained for a dot in a
strong magnetic field. The T--shape DQD close to half--filling has a
similar behavior for small inter--dot coupling $t$, the possibility
of a two-stage Kondo effect leading to a nonmonotonic behavior of
the conductance as function of the gate voltage and magnetic field.
At large inter--dot coupling $t$ the magnetic moments of the two
quantum dots form a ``local" molecular spin--singlet and the
conductance varies monotonically at low temperature. One of the main
results from \cite{co} is the calculation of the conductance $G(T)$
in terms of the spectral density of the active dot A interacting
with the side dot S. The zero temperature conductivity depends on
the total number of electrons in the two dots, i.e., the active dot
A and the side dot S.

Here, we present a $T=0$ K calculation for the dc conductance of the
T-shaped quantum dot system. Our analysis will start from a general
Hamiltonian describing the possible interactions in the double
quantum dot system, i.e., interactions inside each component dot,
inter--dot interactions, and interactions with the reservoirs.
Previous results obtained by Cornaglia and Gempel \cite{co} give the
system's conductance in terms of the total electron density in the
system. However, as we will prove later in the paper, there are
additional contributions related to the inter--dot
electron--electron interaction which were not included in Ref.
\cite{co}. We will also consider the system's conductance in the
presence of a magnetic field whose role is to remove the spin
degeneracy for the possible bound states in the active and side
dots. All our calculations are performed in the $T=0$ K limit, so
finite temperature effects will be neglected. The relevance of
temperature effects due to the different Kondo regimes can be
evaluated by calculating the self energies of the electrons using
the equation of motion method with an appropriate decoupling,
however, the finite temperature conductance of the system will be
the subject of another investigation \cite{cris}.

\section{The Model}

The general hamiltonian of the T--shape double quantum dot
configuration is
\begin{equation}
H=H_{D}+H_{E}+H_{DE}\;.
\end{equation}
Here, $H_D$ describes both the active A and side S dots
\begin{eqnarray}
H_{D}&=&\sum_{i=A,
S}[\epsilon_{i}(n_{i\uparrow}+n_{i\downarrow)}+U_{i}n_{i\uparrow}n_{i\downarrow}]\nonumber\\
&&+t\sum_{\sigma}(d^{\dag}_{A\sigma}d_{S\sigma}+d^{\dag}_{S\sigma}d_{A\sigma})\;,
\end{eqnarray}
where $U_i$ represents the on--site electron--electron Coulomb
interaction and $t$ describes the coupling between the active and
side dots. The operators $d^\dagger_{i\sigma}$ and $d_{i\sigma}$
($i=A, S$) are the standard electron creation and annihilation
operators. The electrons in the left (L) and right (R) electrodes
are described by
\begin{equation}
H_{E}=\sum_{k,\sigma,j}E_{j}c^{\dag}_{ k\sigma j}c_{k\sigma j}\;,
\end{equation}
where the index $j=L,R$; $c^\dagger_{k\sigma j}$ ($c_{k\sigma j}$)
creates (annihilates) an electron with momentum $k$ and spin
$\sigma$ in the $j$ electrode of the configuration. The coupling
between the T--shape DQD and the leads is described by the
Hamiltonian $H_{D-E}$ which has the form:
\begin{equation}
H_{D-E}=\sum_{k,\sigma,j}V_{kj}(d^{\dag}_{A\sigma}c_{k\sigma
j}+c^{\dag}_{k\sigma j}d_{A\sigma})\;.
\end{equation}
All the properties of T-shape DQD configuration can be obtained from
the Green function of the $d$--electrons. The $d$--electron's Green
function can be obtained by different methods including the equation
of motion method (EOM) or the perturbation theory. In the following
we will explore the EOM to extract the electronic Green function and
thereafter the configuration's total conductance.

The properties of the T-shape DQD can be expressed in terms of a
$2\times 2$ Green-function matrix according to the Dyson equation

\begin{equation}
{\bf G^{-1}_{\sigma}(\omega)}= {\bf G^{-1}_{0}(\omega)
 -{\bf \Sigma_{\sigma}(\omega)}}\;,
\end{equation}
where $\bf G_{0}$ is the noninteracting Green function
\begin{equation}
{\bf G^{-1}_{0}(\omega)}= \left(
\begin{array}{cc}
\omega-E_{A}+i\Delta & t\\
t &\omega-E{}_{S} \\
\end{array}
\right)
\end{equation}
with $\Delta = 2\pi N(0)<|V_{kj}|^{2}> $  and
$\Sigma_{\sigma}(\omega)$ is the self--energy matrix as it results
from the Coulomb electron--electron interactions, $U$. In the most
general form the self--energy matrix can be written as
\begin{equation} {\bf \Sigma_{\sigma}(\omega)}= \left(
\begin{array}{ll}
\Sigma_{AA}^{\sigma}(\omega)& \Sigma_{AS}^{\sigma}(\omega)\\
\Sigma_{SA}^{\sigma}(\omega) &\Sigma_{SS}^{\sigma}(\omega) \\
\end{array}
\right)\;,
\end{equation}
a form which accounts both for electron--electron interactions in
each of the two dots and for electron--electron interactions between
the two dots of the configuration. The exact Green's function and
the self--energy of the system satisfy the Luttinger theorem:
\begin{equation}
\int_{-\infty}^{0}d\omega\; \textrm{Tr}\left[\frac{\partial\
{\bf\Sigma_{\sigma}(\omega)}}{\partial\omega}{\bf
G_{\sigma}(\omega)}\right]=0\;,
\end{equation}
where $\textrm{Tr} {\bf A}$ represents the trace of the matrix. The
knowledge of the electronic Green's function permits the calculation
of the total electron density in the system as:
\begin {equation}
n_{d\sigma}=
\textrm{Im}\int_{-\infty}^{0}\frac{d\omega}{\pi}\;\textrm{Tr}\bf
G_{\sigma}(\omega)\;.
\end{equation}
The above expression can be simplified to
\begin{equation}\label{density}
n_{d\sigma}=\frac{1}{\pi}\cot^{-1}\frac{\textrm{Re}\left[
\textrm{det}\bf G^{-1}_{\sigma}(0)\right]}{\textrm{Im}\left[
\textrm{det}\bf G^{-1}_{\sigma}(0)\right]}\;.
\end{equation}

\section{Conductance}

\begin{figure}[b]
\scalebox{0.85}[0.85]{\includegraphics*{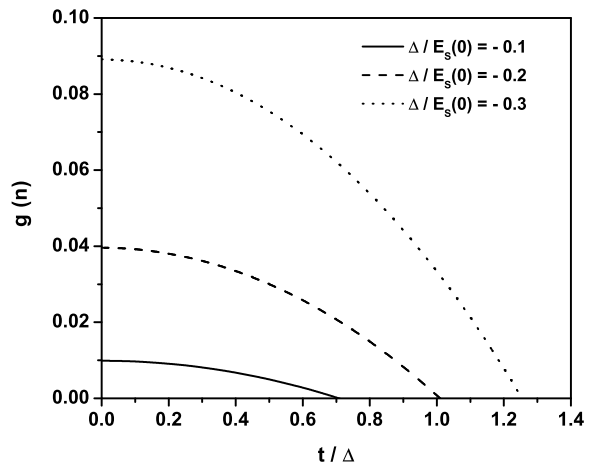}}
\caption{The conductance dependence on the relative coupling between
the active and side quantum dots, $t$, for various values of the
side dot energy level (full line - $\Delta/E_S(0)$=$-0.1$, dashed
line - $\Delta/E_S(0)$=$-0.2$, dotted line - $\Delta/E_S(0)$=$-0.3$)
and fixed energy value for the active dot ($\Delta/E_A(0)=-0.1$).}
\label{figConductance}
\end{figure}

Confinement of electronic systems in small quantum dot
configurations may result in very interesting transport properties.
Here, we calculate the T-shape DQD system's transport properties
following the general formalism introduced by Meir and Wingreen
\cite{meir}. According to Ref. \cite{meir} the current through a
quantum dot system in the presence of an external bias voltage is
given by
\begin{eqnarray}
I&=&\frac{e}{h}\sum_{\sigma}\int
d\omega\left[f(\omega)-f\left(\omega+\frac{eV}{h}\right)\right]\nonumber\\
&&\times\;\;\textrm{Im} \left[\textrm{Tr}\left(\bf \Gamma \bf
G_{\sigma}(\omega)\right)\right]\;,
\end{eqnarray}
where $f(\omega)$ represents the Fermi--Dirac distribution function,
and
\begin{equation}
{\bf \Gamma}= \left(
\begin{array}{ll}
-\Delta & i t\\
i t &0 \\
\end{array}
\right)\;.
\end{equation}
In the zero temperature limit, $T=0$ K, we can evaluate the
conductance of the T-shape DQD configuration ($G=\6I(V)/\6V$) as
\begin{equation}\label{eqG}
G=G_{0}\textrm{Im}\left[\textrm{Tr}({\bf \Gamma
G}_{\sigma}(\omega=0))\right]\;,
\end{equation}
where $G_0=2 \pi e^2/h^2$. The calculation of the system's
conductance as function of the total number of electrons ($n$) is
relatively simple, and a general formula can be given as
\begin{equation}\label{cond1}
g(n)=\f{G}{G_0}=
\frac{\Delta^2E^2_{A}(0)-2t^2E^2_{A}(0)E^2_{S}(0)+2t^4}
{[E^2_{A}(0)E^2_{S}(0)-t^2]^2+\Delta^2E^2_{S}(0)}\;,
\end{equation}
where $E_{A}(0)=E_A-\textrm{Re}\Sigma_{A}(0)$ and
$E_{S}(0)=E_S-\textrm{Re}\Sigma_{S}(0)$ are the renormalized
energies of the bound states in the active, respectively side,
quantum dots of the configuration. Eq. (\ref{cond1}) is an exact
result which shows that the dc conductance of the system depends on
two coupling parameters, $t$ - the coupling between the active and
side dots and $\Delta$ - the coupling between the active dot and the
leads, and the value of the system's self--energy. Accordingly, the
behavior of the system's conductance depends on the selection of the
constituent dots and on the external applied bias. For example, in
Figure \ref{figConductance} we plotted the value of the relative
conductance, $g(n)=G/G_0$, as function of the relative inter dot
coupling $t/\Delta$ for various values of the relative energy level
of the side dot and a fixed value of the relative energy level in
the active dot. Such graphic representations of the T-shape DQD
conductance as function of various interaction energies in the
system allow the optimal selection of the active and side dots. Our
plotting assumes fixed values for the energy levels inside the
active and side dot. This assumption may be questionable as for
many-body effects in the system the initial energy of the bound
states in the two component dots will be changed; however, in most
of the real situations the corrections due to the self-energy on the
value of the two bound states, $E_A$ and $E_S$, are small, and in a
first approximation they can be neglected. Tsvelik and Wiegmann
calculated the self-energy due to the Coulomb interaction $U$
\cite{tsvelik} and proved that the real part of the self-energy
depends linearly on frequency $\omega$. At $T=0$, in the Fermi
liquid approximation, the main contribution to the self-energy comes
from the term $\omega=0$ and accordingly it can be neglected.

On the other hand Eq. (\ref{density}) gives a direct relation
between the system's self energy and the total number of electrons
in the constituent dots. As a result, the conductance of the T-shape
DQD system can be expressed also using the total number of electrons
in the active and side dots. A straightforward calculation, which
implies Eqs. (\ref{density}) and (\ref{cond1}), gives
\begin{equation}
\textrm{Im}\left[\textrm{Tr}({\bf \Gamma
G}_{\sigma}(0))\right]=\sin^{2}(\pi n_{d\sigma})+\frac{t^2}{\Delta
E_{S}(0)}\sin(2\pi n_{d\sigma})\;,
\end{equation}
and in terms of the total number of electrons the system's
conductivity becomes
\begin{equation}\label{cond2}
g(n)=\f{1}{2}\sum_{\sigma}\sin^2\left(\pi
n_{d\sigma}\right)+\frac{t^2}{2\Delta E_{S}(0)}\sum_\sigma\sin(2\pi
n_{d\sigma})\;.
\end{equation}
Note that $E_S(0)$ depends also on the system's self-energy, and
implicitly on the total number of electrons in the system. However,
for the side dot, which is not directly connected to the leads, the
changed of the bound state energy is much smaller that for the
active dot, and such a dependence can be neglected. Eq.
(\ref{cond2}) is an exact result which describe the dependence of
the system's dc conductance on the total number of conduction
electrons. From the experimental point of view the total number of
electrons in the system is controlled using the applied external
bias.

In the absence of an external magnetic field, when the two
considered dots are unpolarized ($n_{d\sigma}$=$n_{d-\sigma}$) the
normalized conductance $g(n)=G/G_{0}$ can be written as
\begin{equation}
g(n)=\sin^2\left(\frac{\pi n}{2}\right)+\frac{t^2}{\Delta
E_{S}(0)}\sin(\pi n)\;,
\end{equation}
where $n=2n_{d\sigma}$ is the total number of electrons in the
active and side dots. It is easy to see that the zero temperature
conductance of the T-shape DQD system vanishes when the total number
of available electrons in the active and side dots is even. This
result was already predicted in Ref. \cite{co}. Our calculation for
the system's conductance also accounts for the intra and inter dots
electron--electron interactions. For a finite occupancy of the
T-shape DQD system, i.e., integer number of electrons occupy the two
possible energy levels, the correction term will cancel ($\sin{(\pi
n)}=0$ for $n$-integer). However, this fact is not an indication
that the electron-electron interactions in the system can be
neglected, the available number of electrons on the active and side
dots being strongly dependent on these interactions and on the
applied, external bias.

A different situation occurs when the degeneracy of the active and
side quantum dots is removed. For example, when a small magnetic
field is applied, the T-shape DQD becomes polarized and the possible
energy levels become spin dependent. In this case, the system's dc
conductance becomes
\begin{eqnarray}
g(n,m)&=&\f{1}{2}\left[1-\cos{(\pi n)}\cos{(\pi
m)}\right]\nonumber\\
&&+\f{t^2}{\Delta E_S(0)}\sin{(\pi n)}\cos{(\pi m)}\;,
\end{eqnarray}
where $m=n_{d\uparrow}-n_{d\downarrow}$ represents the system's
magnetization. For any even number of electrons in the system
($n=0$, $n=2$, and $n=4$), the magnetization value reduces to the
one presented in Ref. \cite{co} for the $n=2$ case, i.e.,
$g(n,m)=\sin^2{(\pi m/2)}$. On the other hand, when the number of
electrons in the system is odd, a different behavior of the
magnetization should be expected, i.e., $g(n,m)=1-\sin^2{(\pi
m/2)}$. Once again, the interaction term in the conductance will
vanish for an integer number of electrons in the system.

\section{Discussions}

In this work we analyzed the conductance of a T-shape DQD. Our main
result is a generalization of the unitary rule of the single-level
Anderson impurity problem for the case of a T-shape DQD . The
system's conductance depends on various interactions inside the
component dots, and on the value of the energy levels in the active
and side dots. Tuning these interactions allow a direct control of
the system conductance. We also presented a calculation of the
system conductance in terms of the occupancy of the two possible
energy levels in the active and side dots. A similar calculation was
also presented in Ref. \cite{co}, however, our result includes
contributions related to the inter-dot interaction, $t$. In the
$t=0$ limit we recover the result presented in Ref. \cite{co}.
Additionally, we proved that such a term can be disregarded when the
energy levels in the T-shape DQD system are occupied by an integer
number of electrons. On the other hand, even at $T=0$ K or at small,
but finite temperatures, when fluctuations are important, the
occupancy of the energy levels can be non-integer, and in such
situations the interaction correction to the system's conductance
becomes important. The differences between our calculation and the
one presented in Ref. \cite{co} are a consequence of how the general
current passing through the system was calculated.  The result
presented by Cornaglia and Grempel \cite{co} only accounts for the
electron density in the active dot, and accordingly the conductance
is calculated considering the inter-dot interaction small, the
system behaving like a single dot with an electron occupancy
$n_{d}=n_{A}+n_{S}$. Such a result will be valid only in the limit
$t^2\ll \Delta E_{S}(0)$.

Sweeping the gate voltage into and trough an odd valley in this
model, corresponds to sweeping the energy of the active dot from
above $\Delta$ to below $-(U+\Delta)$ in the course of which the
total number of electrons $n_{d}$ changes smoothly from $0$ to $2$.
We can obtain three different regimes for the system as function of
various interaction parameters: (i) the ``empty-orbital" when
$n_d=0$ and $G(n_{d})=0$; (ii) the ``mixed-valence" in which the
number of electrons $n_{d}$ begin to increase due to strong charge
fluctuations; and (iii) the ``local moment" regime, in which the
number of electrons $n_{d}$ approaches 1 and the local levels acts
like a local spin in one of the dots, or in both. The latter regime
can give rise to Kondo correlations, and the system has to be
studied at finite temperature.

Our calculation is done in the $T=0$ K limit, however it may be
extended to finite temperatures, $T\ra 0$ K. In this limit, the
imaginary part of the self-energy
\begin{equation}
Im\Sigma (\omega)=[(\omega^2)+(\pi T)^2]/T_{K}\longrightarrow 0\;,
\end{equation}
when $\omega$=$0$ and the real part of the self-energy is constant.
The temperature scale, $T_K$, is set by the Kondo effect and for the
active dot can be calculated following the method proposed in Ref.
\cite{kas}.

The T-shape DQD system is a very promising quantum dot
configuration, both from the fundamental physics and possible
applications point of view. Systems involving single or multiple
quantum dots may provide many opportunities for strong interaction
effects studies. Also they may provide the optimal environment for
the study of the Kondo effect, or of the
Ruderman-Kittel-Kasuya-Yoshida interaction among local spins.
Different other finite temperatures regimes can be considered by
calculating the electronic self-energy using the equation of motion
along with an appropriate decoupling \cite{cris}.


\begin{thebibliography}{99}
\bibitem{anderson} P.W. Anderson, Phys. Rev. {\bf 124} (1961) 41.
\bibitem{singledot} S.M. Cronenwett, T.H. Oosterkamp , and L.P. Kouwenhoven,
Science {\bf 281} (1998) 540.
\bibitem{doubledots} N.J. Craig, J.M. Taylor, E.A. Lester,
C.M. Marcus, M.P. Hanson, and A.C. Gossard,  Science {\bf 304}
(2004) 565.
\bibitem{1} L.G.G.V. Dias da Silva, N.P. Sandler, K. Ingersent, and S.E. Uolla,
Phys. Rev. Lett. {\bf 97} (2006) 096603.
\bibitem{kim} T.--S. Kim and S. Hershfield, Phys. Rev. {\bf 63} (2001) 245326.
\bibitem{tak} K. Takazawa, Y. Imai, and N. Kawkami, J. Phys. Soc.
Japan. {\bf 71} (2002) 2234.
\bibitem{co} P.S. Cornaglia and D.R. Grempel, Phys. Rev. B {\bf 71} (2005) 075305.
\bibitem{zi} R. Zitko and J. Bonca, Phys. Rev. B {\bf 74} (2006) 045312.
\bibitem{wiel}W.G. van der Wiel, S. De Franceschi, J.M. Elzerman, S.
Tarucha, L.P. Kouwenhowen, J. Motohisa, F. Nakajima, and  T. Fukumi,
Phys. Rev. Lett. {\bf 88} (2002) 126803.
\bibitem{meir} Y. Meir and N.S. Wingreen, Phys. Rev. Lett. {\bf
68} (1992) 2512.
\bibitem{kas} V. Kaschcheyevs, A. Aharony, and O. Entin--Wolman ,
Phys. Rev. {\bf 73} (2006) 125338.
\bibitem{cris} M. Crisan, I. Grosu, and I. \c{T}ifrea (in preparation).
\bibitem{tsvelik} A. M. Tsvelik and P. G. Wiegmann, Adv. Phys {\bf
32}, 453 (1983).
\end{thebibliography}
\end{document}